\documentclass[12pt]{article}

\catcode`\@=11
\@addtoreset{equation}{section}

\global\arraycolsep=2pt
\usepackage[dvipdfmx]{graphicx}
\usepackage{mediabb}
\usepackage{amsbsy,amssymb,latexsym,amsfonts,amsmath}
\usepackage{graphicx,color}

\allowdisplaybreaks

\begin{document}

\begin{flushright}
\parbox{4.2cm}
{RUP-21-6}
\end{flushright}

\vspace*{0.7cm}

\begin{center}
{ \Large  Is there supersymmetric Lee-Yang fixed point in three dimensions?}
\vspace*{1.5cm}\\
{Yu Nakayama}
\end{center}
\vspace*{1.0cm}
\begin{center}

Department of Physics, Rikkyo University, Toshima, Tokyo 171-8501, Japan

\vspace{3.8cm}
\end{center}

\begin{abstract}
The supersymmetric Lee-Yang model is arguably the simplest interacting supersymmetric field theory in two dimensions, albeit non-unitary. A natural question is if there is an analogue of supersymmetric Lee-Yang fixed point in higher dimensions. The absence of any $\mathbb{Z}_2$ symmetry (except for fermion numbers) makes it impossible to approach it by using perturbative $\epsilon$ expansions. We find that the truncated conformal bootstrap suggests that candidate fixed points obtained by the dimensional continuation from two dimensions annihilate below three dimensions, implying that there is no supersymmetric Lee-Yang fixed point in three dimensions. We conjecture that the corresponding phase transition, if any, will be the first order transition.
\end{abstract}

\thispagestyle{empty} 

\setcounter{page}{0}

\newpage

\section{Introduction}
The Lee-Yang fixed point is the simplest non-unitary minimal model in two dimensions \cite{Cardy:1985yy}. It has a physical realization as the Ising model under the presence of the critical pure imaginary magnetic  field \cite{Yang:1952be}\cite{Lee:1952ig}. A particular off-critical deformation makes the theory integrable and many exact results such as S-matrix have been computed \cite{Zamolodchikov:1989cf}\cite{Cardy:1989fw}. The Lee-Yang fixed point can be naturally generalized in higher dimensions and we believe that the upper critical dimension is six, where it is described by a Landau-Ginzburg effective field theory with the imaginary potential $V = i\phi^3$ \cite{Fisher:1978pf}.

In two dimensions, there is an $\mathcal{N}=1$ supersymmetric version of the Lee-Yang fixed point realized as the simplest non-unitary $\mathcal{N}=1$ superconformal minimal model \cite{Schoutens:1990vb}\cite{Ahn:1990uq}. It is arguably the simplest interacting $\mathcal{N}=1$ superconformal field theory. A particular off-critical deformation preserves the supersymmetry and it makes the theory integrable \cite{Ahn:1993qa}\cite{Moriconi:1995aj}. Like the Lee-Yang fixed point, many exact results have been computed \cite{Ahn:2000tj}\cite{Kormos:2007qb}. The drawback, however, is that  we do not know the precise physical realization of the supersymmetric Lee-Yang fixed point. We do not know if there is any higher dimensional analogue of the supersymmetric Lee-Yang fixed point, either.

The purpose of this paper is to investigate if there is an analogue of the supersymmetric Lee-Yang fixed point in higher dimensions. Since the number of supersymmetry preserved here is half of the smallest supersymmetry in four dimensions, our main target will be three dimensions, but we study the question in terms of the dimensional continuation. The first question is if we can propose a simple effective field theory description that can be used to do perturbative $\epsilon$ expansions. It turns out that the supersymmetric Lee-Yang fixed point does not possess any $\mathbb{Z}_2$ symmetry (except for the fermion numbers), but we find that any perturbative approaches based on $\epsilon$ expansions preserve the $\mathbb{Z}_2$ symmetry. As a consequence, we  find that the supersymmetric Lee-Yang fixed point cannot be accessed by $\epsilon$ expansions (unlike  the Lee-Yang fixed point near six dimensions).

As a non-perturbative tool to investigate a non-unitary fixed point, we use the  truncated conformal bootstrap proposed in  \cite{Gliozzi:2013ysa}\cite{Gliozzi:2014jsa}. The idea is to try to solve the conformal bootstrap equation by a relatively small number of operators. One cannot solve the conformal bootstrap equation exactly with finitely truncated operators, but demanding that it is approximately solved in the vicinity of the crossing symmetric point gives a constraint on the truncated spectrum. While this is not a controlled approximation,  it seems to give a qualitative (and sometimes quantitative) picture of the non-trivial fixed points by changing space-time dimensions. 

We use the truncated conformal bootstrap to study the fate of the supersymmetric Lee-Yang fixed point by changing the space-time dimensions from two while preserving the structure of the operator product expansions. We will see that the candidate fixed points annihilate before reaching three dimensions. It indicates that the supersymmetric Lee-Yang fixed point may not exist in three dimensions. The absence of the fixed point means that the corresponding phase transition, if any, will be the first order transition rather than the second order transition (the latter being the case in the ordinary Lee-Yang model).

\section{Supersymmetric Lee-Yang fixed point}

\subsection{As a (super)conformal minimal model in two dimensions}
The supersymmetric Lee-Yang fixed point in two dimensions is defined by the superconformal minimal model $SM_{2,8}$. It has the central charge $c=-\frac{21}{4}$ and hence it is non-unitary. The supersymmetric Kac table is shown in Table \ref{tb1}. Superconformal primary operators with $h = 0, \pm\frac{1}{4}$ are in the Neveu-Schwarz sector and superconformal primary operators with $h= -\frac{3}{32},-\frac{7}{32}$ are in the Ramond sector.  We will not be interested in the Ramond sector in this paper because there is no analogue in higher dimensions.

\begin{table}[htb]
\begin{center}
\begin{tabular}[t]{|c|c|c|c|c|c|c|c|}
\hline
$h$ & $0$ & $-\frac{3}{32}$  & $-\frac{1}{4}$ & $-\frac{7}{32}$ & $-\frac{1}{4}$ & $-\frac{3}{32}$ & $0$ \\ \hline
\end{tabular}
 \caption{Supersymmetric Kac table of $SM_{2,8}$.}
  \label{tb1}
\end{center}
\end{table}

Remarkably, the supersymmetric Lee-Yang fixed point is also described by a fermionic version of the conformal minimal model $M_{3,8}$ \cite{Schoutens:1990vb}\cite{Melzer:1994qp}. It has the Kac table shown in Table \ref{tb2}. The $\mathbb{Z}_2$ symmetry of the $M_{3,8}$ minimal model is identified as the fermion number. The relation to the supersymmetric Lee-Yang fixed point is an example of the generalized Jordan-Wigner duality of $\mathbb{Z}_2$ symmetric minimal models as studied in \cite{Runkel:2020zgg}\cite{Hsieh:2020uwb}.

\begin{table}[htb]
\begin{center}
\begin{tabular}[t]{|c|c|c|c|c|c|c|c|}
\hline
$h$ & $0$ & $-\frac{7}{32}$  & $-\frac{1}{4}$ & $-\frac{3}{32}$ & $\frac{1}{4}$ & $\frac{25}{32}$ & $\frac{3}{2}$ \\ \hline
$h$ & $\frac{3}{2}$ & $\frac{25}{32}$  & $\frac{1}{4}$ & $-\frac{3}{32}$ & $-\frac{1}{4}$ & $-\frac{7}{32}$ & $0$ \\ \hline
\end{tabular}
 \caption{ Kac table of $M_{3,8}$.}
  \label{tb2}
\end{center}
\end{table}

Let us denote the bosonic primary operator with the lowest conformal dimension by $\phi$, whose conformal dimension is given by $\Delta_\phi = h + \bar{h} = -\frac{1}{2}$, and study the Virasoro fusion rule of $\phi \times \phi $. It is given by
\begin{align}
[\phi] \times [\phi] = [1] + [\phi] + [\phi^2] \ , \label{fusion}
\end{align}
where $\phi^2$ is a Virasoro primary operator whose conformal dimension $\Delta_{\phi^2} = \frac{1}{2}$. As a superconformal minimal model, it is identified as $G_{-\frac{1}{2}}\bar{G}_{-\frac{1}{2}} \phi$ (here $G_{-\frac{1}{2}}$ is the superysmmetry generator). It is the existence of $[\phi]$ and $[G_{-\frac{1}{2}}\bar{G}_{-\frac{1}{2}} \phi]$  in the $[\phi] \times [\phi]$ fusion rule what characterizes the supersymmetric Lee-Yang fixed point.

Because of this fusion rule, the supersymmetric Lee-Yang fixed point does not possess any extra $\mathbb{Z}_2$ symmetry (except for the left-right fermion numbers). It should be contrasted with the supersymmetric Ising model (or fermionic version of the tri-critical Ising model) with the extra $\mathbb{Z}_2$ symmetries (beyond fermion numbers). The appearance of $[\phi]$ in $[\phi] \times [\phi]$ is the reason why we call it supersymmetric ``Lee-Yang". The supersymmetric Lee-Yang fixed point has two relevant operators (other than the identity). One of them preserves supersymmetry and results in a massive scaling theory. The other particular non-supersymmetric combination is a critical deformation, which results in a free massless fermion in the infrared. 

There is one point in the spectrum that makes the supersymmetric Lee-Yang fixed point slightly different from the Lee-Yang fixed point. In the Lee-Yang fixed point, the Virasoro conformal family of the scalar operator with the lowest conformal dimension $\phi$ (with $\Delta_{\phi} = -\frac{2}{5}$) contains a singular vector at level two, and the  (quasi-primary) spin two operator with the lowest conformal dimension in the theory is $L_{-2} 1$ with $\Delta_{2}= 2$ (i.e. the energy-momentum tensor), rather than $L_{-2}\phi$, which is identified with $L_{-1}^2 \phi$ from the null vector condition. In contrast, in the supersymmetric Lee-Yang fixed point, the spin two operator with the lowest conformal dimension is $L_{-2}\phi$ with $\Delta_{2}= \frac{3}{2}$ because it is not a singular vector. 

\subsection{Lagrangian descriptions?}
To address the question if there is an analogue of the supersymmetric Lee-Yang fixed point in three dimensions, let us first investigate a possibility to realize it as an effective field theory of the Landau-Ginzburg type.
General $\mathcal{N}=1$ supersymmetric Landau-Ginzburg effective field theories in two and three dimensions are described by a superpotential $W(\Phi)$. Here $\Phi = \phi + \theta \psi + \cdots$ is a (real) superfield whose bottom component is a real scalar $\phi$. The fermion $\psi$ is Majorana in both two and three dimensions.

 In terms of the superpotential, the Landau-Ginzburg effective action is given by 
\begin{align}
S = \int d^dx \left( \frac{1}{2}\partial^\mu \phi \partial_\mu \phi + \frac{i}{2} \bar{\psi} \partial^\mu \gamma_\mu \psi +\frac{1}{2} W''(\phi) \bar{\psi} \psi -\frac{1}{2} W'(\phi)^2 \right) .
\end{align}
We first argue that the supersymmetric Lee-Yang fixed point in two dimensions and three dimensions (if any for the latter) cannot be described by a monomial superpotential that would usually become a starting point in perturbative $\epsilon$ expansions. 

The reason why the monomial superpotential such as $W = \Phi^3$ or $\Phi^4$ does not work for our purpose is that theories with a monomial superpotential have extra $\mathbb{Z}_2$ symmetry. If the superpotential is an even  monomial, we have an ordinary $\mathbb{Z}_2$ symmetry, under which $\phi \to - \phi$ (and $\bar{\psi}\psi \to  \bar{\psi}\psi$). If the superpotential is an odd monomial, we have a chiral (in two dimensions)  or time-reversal (in three dimensions) $\mathbb{Z}_2$ symmetry, under which $\phi \to - \phi$ and $\bar{\psi}\psi \to - \bar{\psi}\psi$. In both cases, they cannot describe the supersymmetric Lee-Yang fixed point with no $\mathbb{Z}_2$ symmetry.\footnote{A possible loophole in the discussion here is when the $\mathbb{Z}_2$ symmetry is spontaneously broken. In the chiral case, we need to find a non-trivial topological field theory to accommodate the anomaly. In any case, the fixed point studied in the $\epsilon$ expansions cannot see this effect.}

If the superpotential is given by a  polynomial, e.g. $W = g_2 \Phi^2 + i g_3 \Phi^3  + g_4\Phi^4$, it is consistent with the assumed absence of the $\mathbb{Z}_2$ symmetry, but it cannot be used to study the fixed point perturbatively in $\epsilon$ expansions. Indeed, if we work near four dimensions, the would-be fixed point accessible in $\epsilon$ expansions is $g_3= O(\sqrt{\epsilon})$ (with $g_2 = g_4=0$) in $d=4-\epsilon$ dimensions and it is ``$\mathbb{Z}_2$ symmetric".\footnote{Unless we are in two or three dimensions, there is no natural $\mathbb{Z}_2$ transformation that flips the sign of $\bar{\psi}\psi$. The one-loop renormalization group equation, however, is $\mathbb{Z}_2$ symmetric.} The actual one-loop computation by dimensional continutation indicates that the only available fixed point is a ``unitary" one \cite{Fei:2016sgs}\cite{Gies:2017tod} (which is believed to describe the supersymmetric Ising model in three dimensions with an Hermitian superpotential \cite{Grover:2013rc}\cite{Bashkirov:2013vya}\cite{Iliesiu:2015qra}\cite{Shimada:2015gda}), so the study of the supersymmetric Lee-Yang fixed point based on the $\epsilon$ expansion seems quite challenging. 

It is likely that the supersymmetric Lee-Yang fixed point is non-unitary but a ``real" conformal field theory in the sense discussed in \cite{Gorbenko:2018ncu}. The same structure is sometimes called PT-symmetric in other literature. From the Landau-Ginzburg effective field theory viewpoint, it is described by the superpotential whose coefficient is pure imaginary for the odd powers of $\Phi$ and real for the even powers of $\Phi$. The superpotential is not Hermitian if the odd powers are present, but it has a non-trivial PT-symmetry $\bar{W}(\Phi) = W(-\Phi)$. In other words, $W$ is real analytic as a function of $i\Phi$. 

\section{Truncated conformal bootstrap}
As we have seen, it seems difficult to approach the supersymmetric Lee-Yang fixed point by using conventional $\epsilon$ expansions. We need to pursue a non-perturbative method. In this section,  we use the truncated conformal bootstrap with dimensional continuation to investigate the possibility to find a supersymmetric Lee-Yang fixed point in higher dimensions. Our main assumption, which we simply assume without proof, is that the supersymmetric Lee-Yang fixed point preserves conformal symmetry in higher dimensions as well.
\subsection{General methods}

In the conformal bootstrap, we are interested in conformal invariant four-point functions. More specifically, in this paper, we focus on the four-point functions of identical scalars $\langle \phi(0) \phi(1) \phi(z) \phi(\infty) \rangle$.
Let us introduce the operator product expansion\footnote{Hereafter, the square bracket denotes the global conformal family rather than Virasoro conformal family. Accordingly, what we mean by the conformal block is a global conformal block rather than the Virasoro conformal block.}
\begin{align}
[\phi] \times [\phi] = \sum_i c_{\phi\phi i} [O_i]
\end{align}
 to compute the four-point functions in two different channels by exchanging $z$ and $1-z$. 
By using the conformal block $G_{\Delta,l}$ (our normalization is taken from \cite{ElShowk:2012ht}), the resultant crossing equation is given by
\begin{align}
\sum_{O_i \neq 1} c_{\phi\phi i}^2 \frac{v^{\Delta_\phi} G_{\Delta_i,l_i}(u,v) - u^{\Delta_{\phi}} G_{\Delta_i,l_i}(v,u)}{u^{\Delta_\phi} - v^{\Delta_{\phi}}} = 1 \ . \label{crossing}
\end{align}
Here $u= z\bar{z}$ and $v = (1-z)(1-\bar{z})$. This is the starting point of all the conformal bootstrap approaches to conformal field theories in any dimensions (see e.g. \cite{Poland:2018epd} for a review).

This is an infinite dimensional constraint as a function of $u$ and $v$. We will be interested in the constraint only from the vicinity of the crossing symmetric point $u=v$, where the convergence of the conformal block expansion is rapid. For this purpose, we introduce new variables $a$ and $b$ by $u = \frac{a^2-b}{4}$, $v= 1-a + \frac{a^2-b}{4}$, and expand \eqref{crossing} in terms of $a-1$ and $b$ around $a=1$ and $b=0$. This gives a set of equations that are labeled by two semi-positive integers $m$ and $n$ (such that $m+n \neq 0$):
\begin{align}
\sum_{O_i \neq 1} c_{\phi \phi i}^2 f_{\Delta_\phi,\Delta_i}^{(2m,n)} = 0 \label{homo}
\end{align}
where
\begin{align}
f_{\alpha,\beta}^{(2m,n)} = \left(\partial_a^{2m} \partial_b^n \frac{v^\alpha G_{\beta}(u,v) - u^\alpha G_{\beta}(v,u)}{u^\alpha -v^\alpha} \right)|_{a=1,b=0} \ . 
\end{align}
One can compute $f_{\alpha,\beta}^{(2m,n)}$  numerically by using the recursion relation obtained in \cite{ElShowk:2012ht}. Note that only even powers of $a-1$ appear in the crossing equation of identical scalar operators. 

In the truncated conformal bootstrap, we first truncate the spectrum and take only a finite number of terms in the operator product expansions. At the same time, we demand that sufficiently many homogeneous equations \eqref{homo} are satisfied. Then the parameters in the spectrum (e.g. conformal dimensions of operators) must be constrained, and we may be able to make non-trivial predictions of the operator spectrum. In \cite{Gliozzi:2013ysa}\cite{Gliozzi:2014jsa}, they declare if this happens to be the case, the conformal field theory is truncatable. At this point, it is not apriori obvious if the conformal field theory of interest is truncatable, but they have demonstrated that Lee-Yang fixed point and critical Ising model may be approached in this way (see also \cite{Gliozzi:2015qsa}\cite{Nakayama:2016cim}\cite{Gliozzi:2016cmg}\cite{Hikami:2017hwv}\cite{Hikami:2017sbg}\cite{Li:2017ukc}\cite{Hikami:2018mrf}\cite{Rong:2020gbi} for the other examples that the method has been applied).

To be more specific, let us assume that equations \eqref{homo} are satisfied with non-zero $c_{\phi \phi i}^2$ with $N$ operators in the operator product expansion. It is convenient to regard them as linear homogeneous equations, $c_{\phi\phi i}^2$ being a vector with $i$ index and $f^{(2m,n)}_{\Delta_\phi,\Delta_i}$ being a matrix (with $i$ index summed). Then we demand minors of order $N$ constructed out of $f^{(2m,n)}_{\Delta_\phi,\Delta_i}$ must vanish. We may use these vanishing conditions to make a non-trivial predictions on the conformal dimensions of operators that appear in $f^{(2m,n)}_{\Delta_\phi,\Delta_i}$ through the truncated spectrum.

Since the vanishing conditions depend on the choice of the minors, we have to specify them. We use the same dictionary used in \cite{Gliozzi:2013ysa}. Corresponding to the set of integers $(m,n)$ in $f^{(2m,n)}_{\Delta_\phi,\Delta_i}$, we assign another integer label:
\begin{align}
1,2,3,4,5,6 \leftrightarrow (1,0),(2,0),(0,1),(0,2),(1,1) 
\end{align}
For example with $N=2$, $d_{12} = 0$ means that we demand minors constructed out of $f^{(2,1)}_{\Delta_\phi,\Delta_i}$ and $f^{(4,0)}_{\Delta_\phi,\Delta_i}$ vanishes:
\begin{align}
\det \begin{pmatrix}
f^{(2,0)}_{\Delta_{\phi},\Delta_1} & f^{(4,0)}_{\Delta_{\phi},\Delta_1} \\
f^{(2,0)}_{\Delta_{\phi}, \Delta_2} & f^{(4,0)}_{\Delta_{\phi}, \Delta_2} \\
\end{pmatrix}
 =  0  \ , 
\end{align}
which gives an equation among $\Delta_{\phi}$, $\Delta_1$ and $\Delta_2$. More technical details on the truncated conformal bootstrap can be found in the literature.

Given the rapid convergence of the conformal block expansions, it is reasonable to keep the operators  with the lower conformal dimensions for each spin. Inclusions of extra operators make the size of the determinant larger and it becomes more difficult to study their zeros in numerics. In unitary conformal field theories, the spin two operator with the lowest conformal dimension is always the energy-momentum tensor and the conformal dimension is fixed to be $\Delta_{2} = d$. In non-unitary conformal field theories, it is not automatically guaranteed. However, in the Lee-Yang fixed point, it is also the case at least in two dimensions, and this property was implicitly employed in the study of \cite{Gliozzi:2013ysa}\cite{Gliozzi:2014jsa}. As we have seen, this is not the case in the supersymmetric Lee-Yang fixed point and we have to determine the conformal dimensions of the lowest spin two operator as well. This makes it more non-trivial to find a reasonable truncation and constraints.

\subsection{Supersymmetric Lee-Yang fixed points}
Now we would like to study the truncated conformal bootstrap for the supersymmetric Lee-Yang fixed point. We need to truncate the fusion rule of the supersymmetric Lee-Yang fixed point \eqref{fusion} in such a way that it can be generalized in higher dimensions than two.

Our first trial is to truncate the operator product expansion in the minimal way as
\begin{align}
\phi\times \phi = 1+ [\phi] + [\phi^2] + [O_2] \ .
\end{align}
Here, $O_2$ is the spin two operator with the lowest conformal dimension $\Delta_2$. As we have mentioned, $\Delta_2$ may be smaller than the space-time dimensions $d$. We further assume the supersymmetry relation $\Delta_{\phi^2} = \Delta_{\phi} + 1$, which will be the single non-trivial information we can use to characterize the supersymmetric Lee-Yang fixed point. Without this, we could not distinguish it from the truncated operator product expansions in the Lee-Yang fixed point.

We may choose two out of various minors to determine two free parameters $\Delta_{\phi}$ and $\Delta_2$. There are some arbitrariness to choose them, and our choice is very heuristic: we simply choose the two minors whose zero give the closest conformal dimensions that we do know in two dimensions as the supsereymmetric Lee-Yang fixed point (i.e. $\Delta_{\phi} = -\frac{1}{2}$ and $\Delta_{2}= \frac{3}{2})$. Then we change the space-time dimensions $d$ to see what will happen. With this strategy, we have determined to show the results from the vanishing minors of $d_{123} = 0$ and $d_{124} =0$.

\begin{figure}[htbp]
	\begin{center}
		\includegraphics[width=12.0cm,clip]{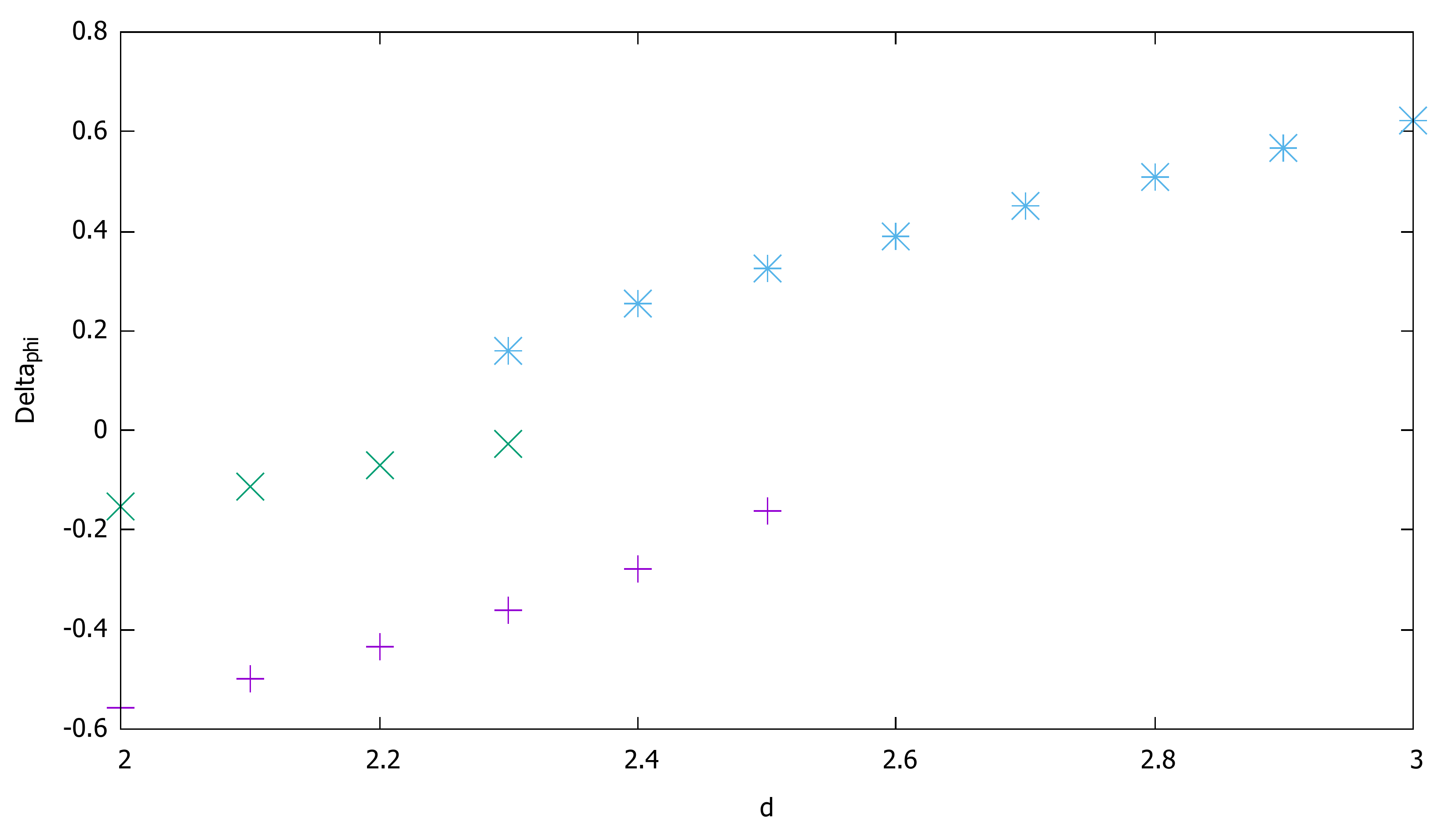}
	\end{center}
	\caption{$\Delta_{\phi}$ of Supersymmetric Lee-Yang fixed point as a function of the space-time dimensions $d$.}
	\label{fig:1-1}
\end{figure}

\begin{figure}[htbp]
	\begin{center}
		\includegraphics[width=12.0cm,clip]{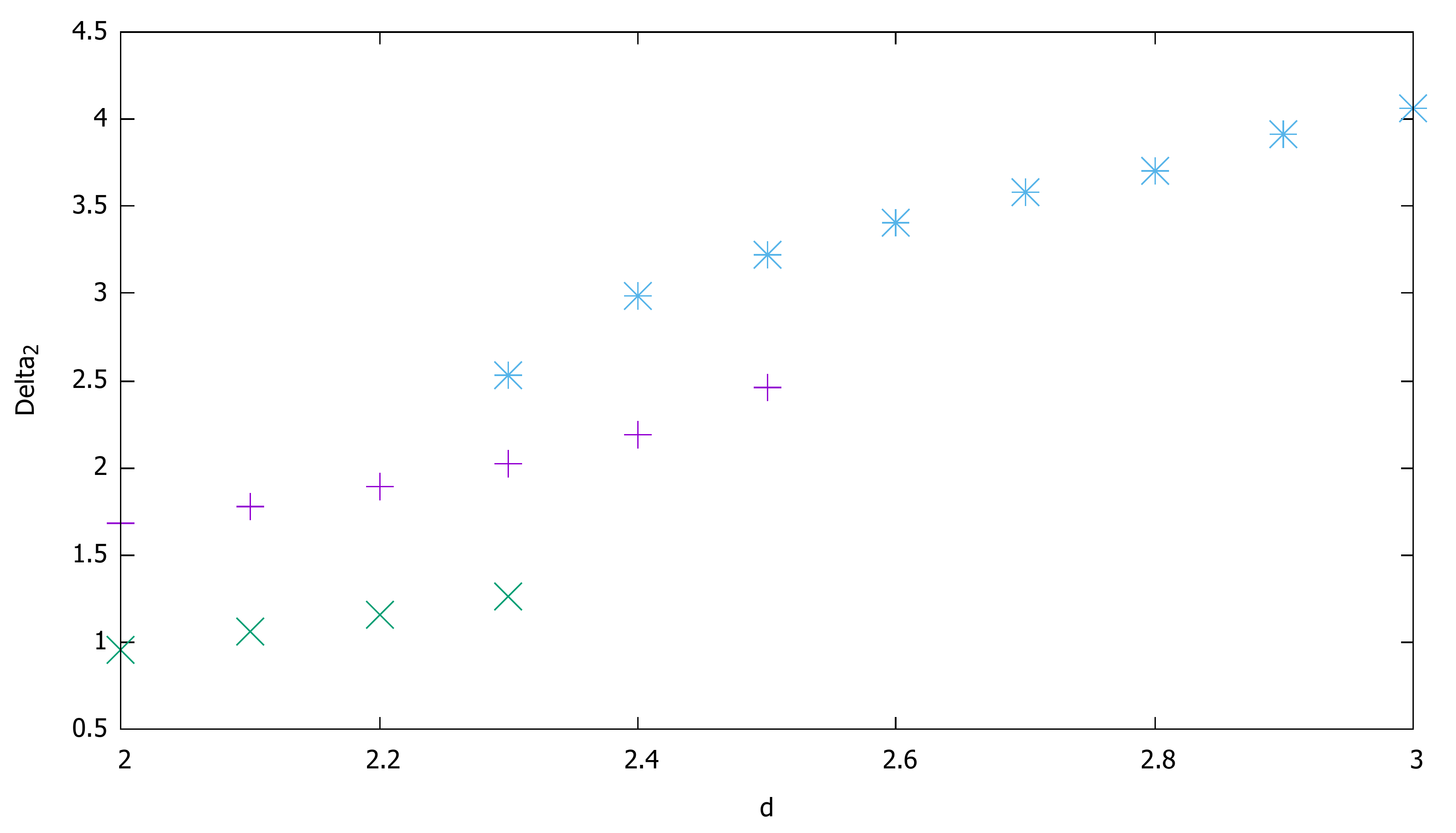}
	\end{center}
	\caption{$\Delta_2$ of Supersymmetric Lee-Yang fixed point as a function of the space-time dimensions $d$.}
	\label{fig:1-2}
\end{figure}

The numerical solution by dimensional continuation is shown in Fig \ref{fig:1-1} and Fig  \ref{fig:1-2}.  Let us focus on the purple crosses. A candidate solution in two dimensions (with $\Delta_\phi = -0.56$ and $\Delta_{2} = 1.68$) disappears above $d=2.5$ dimensions, around which $\Delta_{2}$ reaches the unitarity bound $\Delta_{2} = d$. We can see there are other series of solutions, but none of them can be continued to two dimensions or three dimensions.


In this truncation, we have not assumed the existence of the energy-momentum tensor in the operator product expansion, and the disappearance of the fixed point may be related to the possibility that beyond $d=2.5$, the energy-momentum tensor becomes the most relevant spin two operator. 

To address this possibility, we have added more operators in the truncation. The second trial is to truncate the operator product expansion as
\begin{align}
\phi \times \phi =1+ [\phi] + [\phi^2] + [O_2] +[T] + [O_4] \ , 
\end{align}
Here $T$ is the energy-momentum tensor with $\Delta_T=d$, and $O_4$ is the spin four operator with the lowest conformal dimension. We assume the supersymmetry relation $\Delta_{\phi^2} = \Delta_{\phi} + 1$. Furthermore, to avoid the technical issue,\footnote{As we can immediately see, if $\Delta_2$ is a free parameter, all the minors always become zero when $\Delta_2 =d$ (for any other parameters). The existence of huge  regions of the trivial solutions would spoil our numerical search for non-trivial zeros, so we want to introduce the condition that forbids the solution $\Delta_2 = d$.}  we make the (biased) ansatz $\Delta_{2} = \Delta_{\phi} + 2$, which is true in $d=2$ dimensions but is not theoretically guaranteed in $d>2$. In this paper, with these ansatz we study the spectrum obtained by vanishing minors of $d_{12346} = 0$ and $d_{12456}=0$ by dimensional continuation. 

\begin{figure}[htbp]
	\begin{center}
		\includegraphics[width=12.0cm,clip]{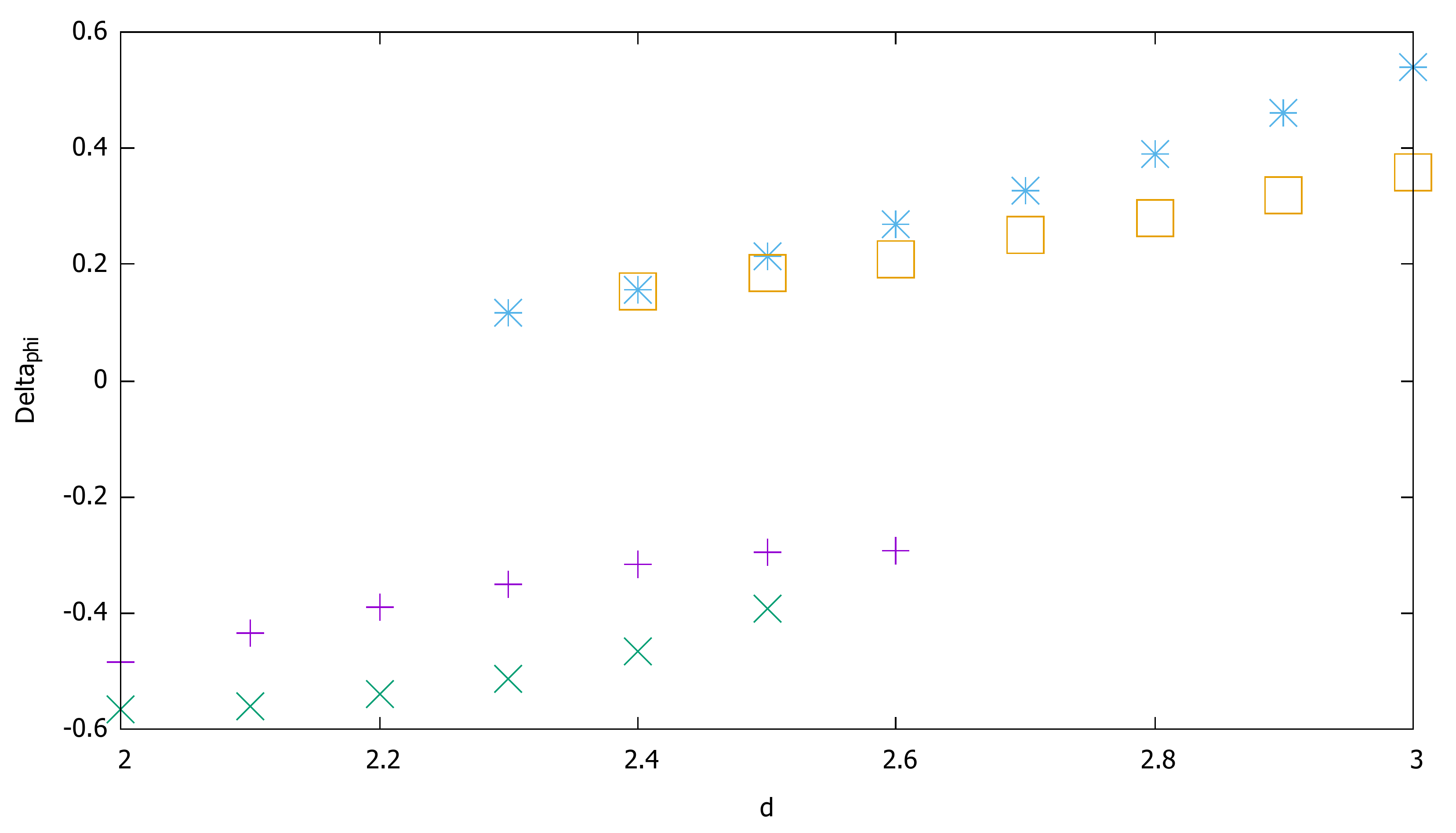}
	\end{center}
	\caption{ $\Delta_{\phi}$ of Supersymmetric Lee-Yang fixed point as a function of the space-time dimensions $d$.}
	\label{fig:2-1}
\end{figure}

\begin{figure}[htbp]
	\begin{center}
		\includegraphics[width=12.0cm,clip]{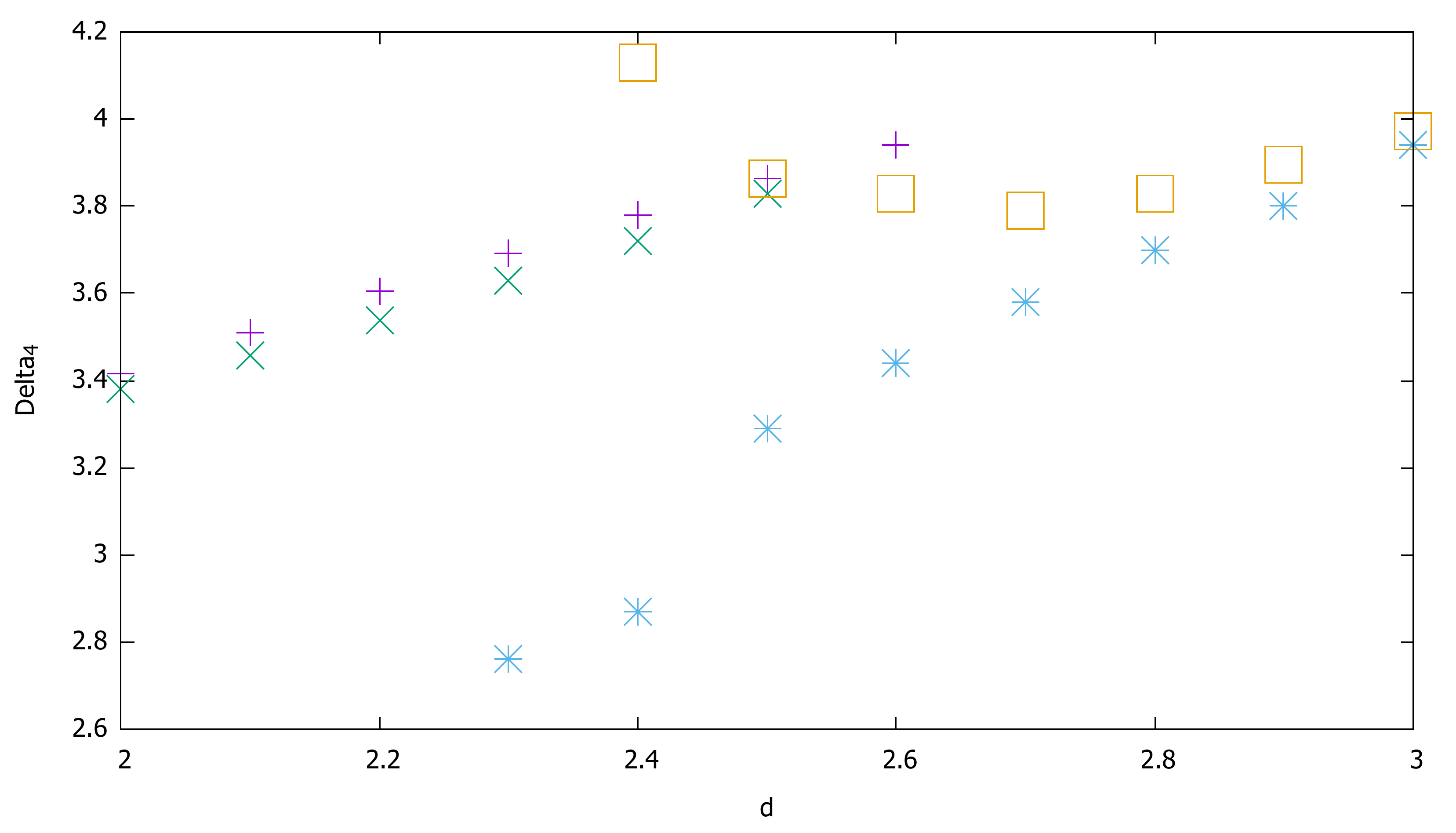}
	\end{center}
	\caption{$\Delta_4$ of Supersymmetric Lee-Yang fixed point as a function of the space-time dimensions $d$.}
	\label{fig:2-2}
\end{figure}

The numerical solution by dimensional continuation is presented in Fig \ref{fig:2-1} and Fig \ref{fig:2-2}. Candidate solutions in two dimensions (a purple cross with $\Delta_\phi = -0.485$ and $\Delta_4 =3.42$) again disappear around $d = 2.6$ dimensions. The way they disappear is more interesting: two solutions annihilate each other, and probably they disappear into complex solutions. This is reasonable because after all we are solving algebraic equations for $\Delta_\phi$ and $\Delta_4$ and when we change the external parameters (i.e. space-time dimensions $d$ here), the real solutions can annihilate into complex solutions.


Conceptually, the annihilation of solutions of crossing equations is similar to what they have observed in the solutions of renormalization group fixed point equations \cite{Kaplan:2009kr}\cite{Gorbenko:2018ncu}. There, they claim that the annihilation into complex fixed point of the renormalization group may be associated with the walking behavior in the first order phase transition. We do not know if this argument also applies to a solution of the conformal bootstrap equation, but aside from the walking behavior, we conjecture that the corresponding phase transition, if any, will be the first order transition in three dimensions.

\subsection{Other fixed points as a benchmark}

In the previous sections, we have seen that the dimensional continuation of the supersymmetric Lee-Yang fixed points obtained in the truncated conformal bootstrap seems to annihilate before reaching three dimensions. In order to see if this is a non-trivial phenomenon, we would like to compare it with what happens in the other fixed points that we may find in the truncated conformal bootstrap  by dimensional continuation (with different assumptions on the spectrum). The following examples can been seen as benchmark tests for the dimensional continuation of the truncated conformal bootstrap.

Let us first study the dimensional continuation of the Lee-Yang fixed point.
The truncation we study as a benchmark is 
\begin{align}
\phi \times \phi = 1 +  [\phi] + [T] + [O_4] \ , 
\end{align}
where $\Delta_{\phi}$ and $\Delta_{4}$ are parameters to be determined. In two dimensions, we know they are given by $\Delta_{\phi} = -\frac{4}{5}$ and $\Delta_{4} = 4$.

\begin{figure}[htbp]
	\begin{center}
		\includegraphics[width=12.0cm,clip]{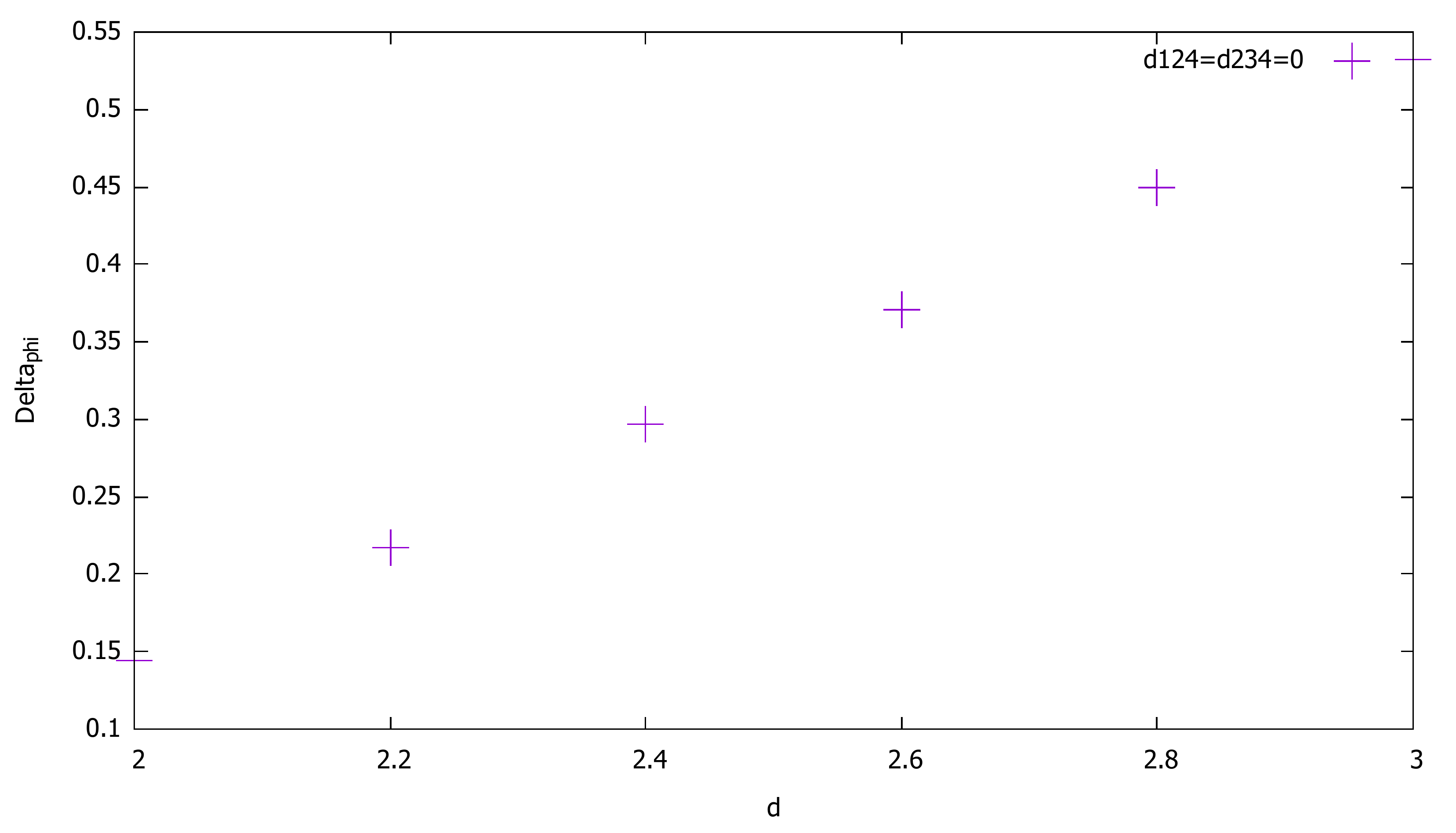}
	\end{center}
	\caption{$\Delta_{\phi}$ of the supersymmetric Ising model as a function of the space-time dimensions $d$.}
	\label{fig:4-1}
\end{figure}

\begin{figure}[htbp]
	\begin{center}
		\includegraphics[width=12.0cm,clip]{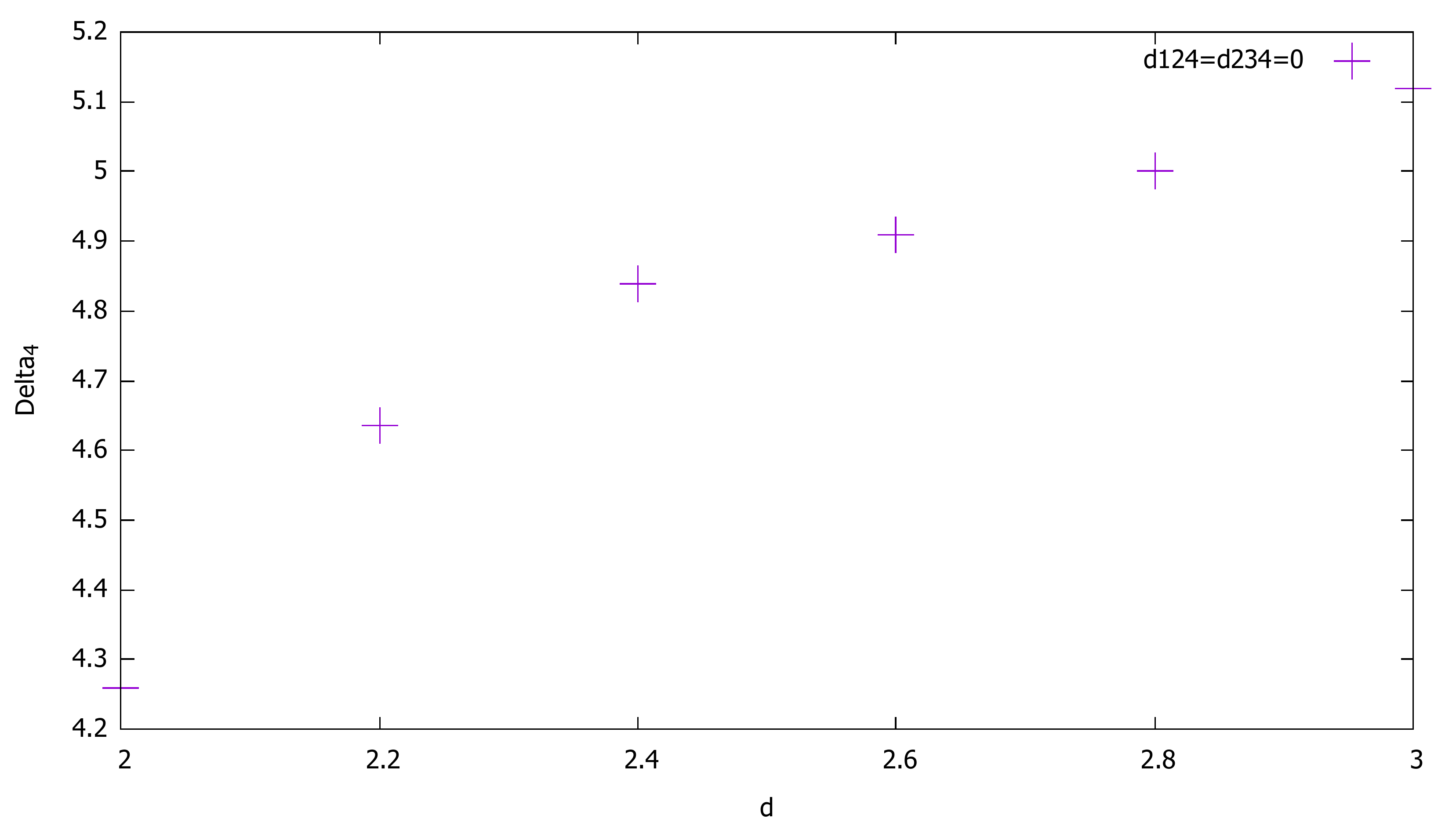}
	\end{center}
	\caption{$\Delta_4$ of the supersymmetric Ising model as a function of the space-time dimensions $d$.}
	\label{fig:4-2}
\end{figure}

We can study the prediction for $\Delta_\phi$ and $\Delta_4$ from the vanishing of $d_{123} = d_{234} = 0 $ by changing the     space-time dimensions. The choice of the vanishing minors is the same one used in \cite{Gliozzi:2013ysa}. 
The results are shown in Fig \ref{fig:4-1} and Fig \ref{fig:4-2}. As we see, the candidate fixed points  (purple crosses) do not disappear and the dimensional continuation predicts a non-trivial fixed point in  three dimensions. In particular, one may be able to pass $\Delta_\phi = 0$. This is in stark contrast with the case studied in the supersymmetric Lee-Yang fixed point above. The dimension at which $\Delta_{\phi}=0$ (here $d\sim 2.6$) has been studied in \cite{Hikami:2017hwv} as a ``critical dimension". Our prediction of the conformal dimension of $\Delta_{\phi} = 0.213$ in three dimension is same as in  \cite{Gliozzi:2013ysa} (so our numerical implementation is compatible with theirs). The latest predictions from the other method can be found in \cite{1854466}.

We may further continue the sace-time dimensions until we reach $d=6$, where it is given by a free field theory with conformal dimensions $\Delta_\phi = 2.0$ and $\Delta_4=8.0$. We can further continue the space-time dimensions above six dimensions, but the formal solution in $d>6$ dimensions is not immediately clear (see, however, \cite{Rong:2020gbi}).\footnote{At some point, we find that $C_T$ becomes negative within this truncation. This was observed by the author back in 2015.}

\begin{figure}[htbp]
	\begin{center}
		\includegraphics[width=12.0cm,clip]{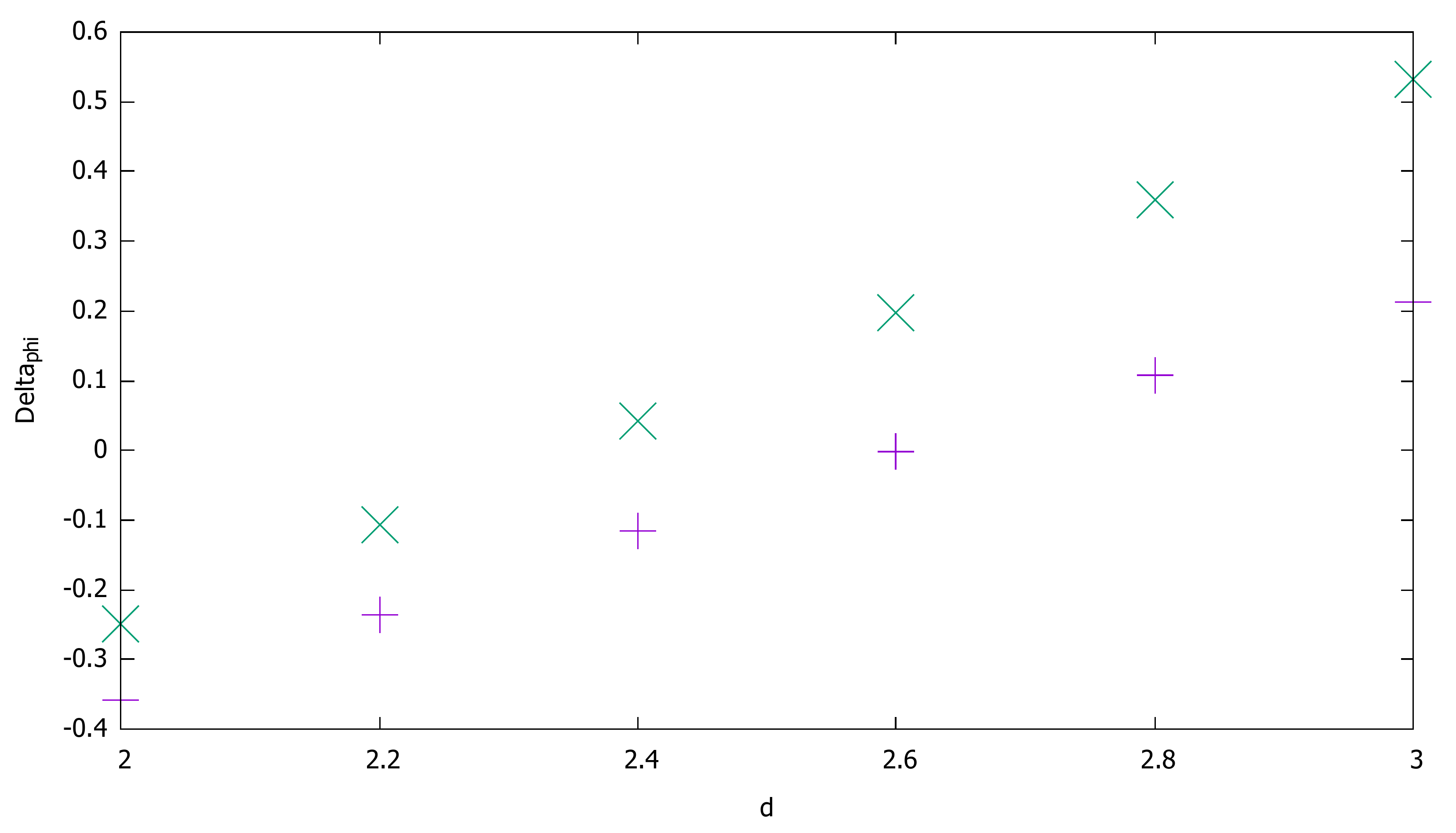}
	\end{center}
	\caption{$\Delta_{\phi}$ of Lee-Yang fixed point as a function of the space-time dimensions $d$.}
	\label{fig:3-1}
\end{figure}
\begin{figure}[htbp]
	\begin{center}
		\includegraphics[width=12.0cm,clip]{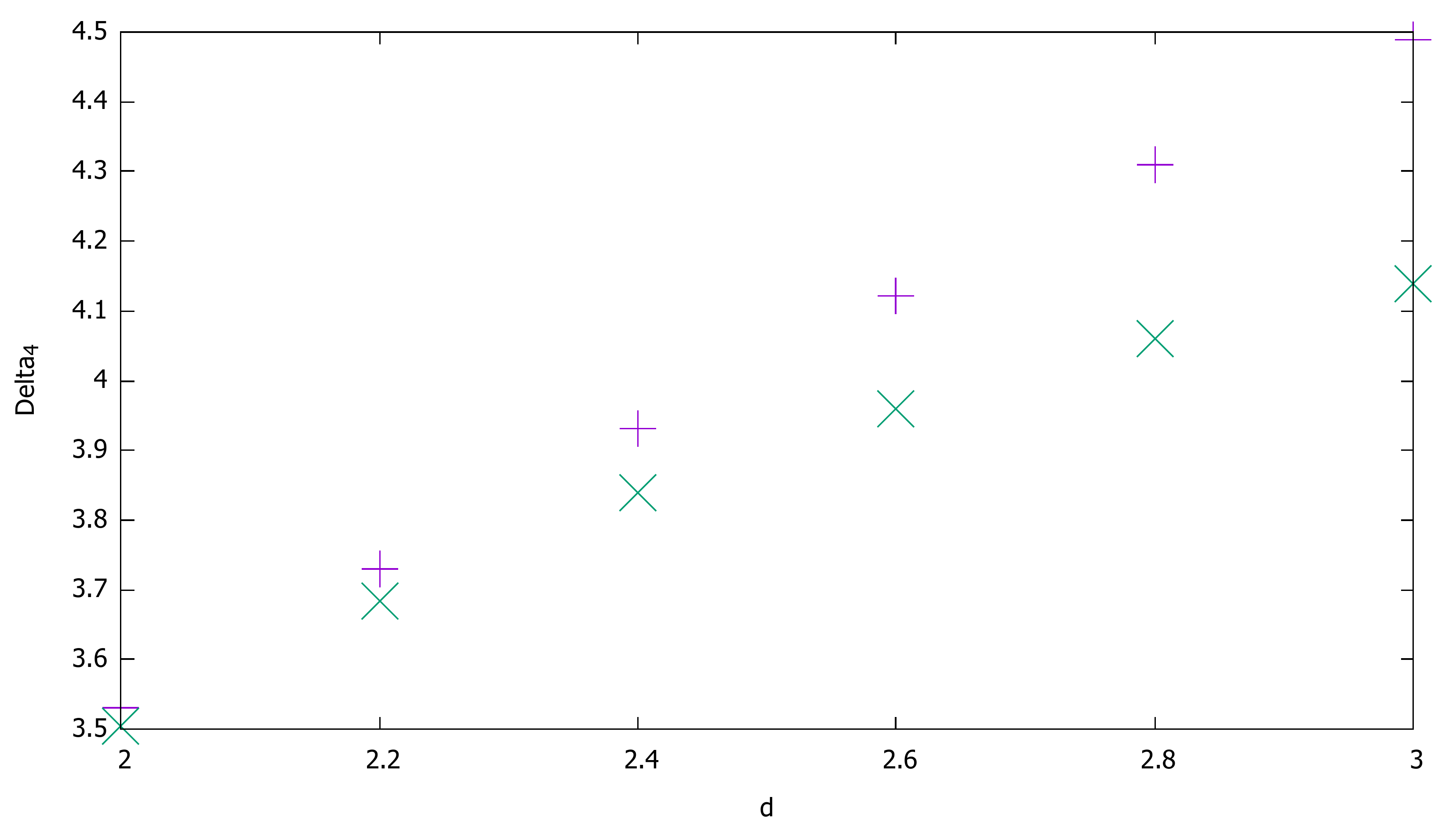}
	\end{center}
	\caption{$\Delta_4$ of Lee-Yang fixed point as a function of the space-time dimensions $d$.}
	\label{fig:3-2}
\end{figure}

Another interesting class is the $\mathcal{N}=1$ supersymmetric Ising model, which is described by $SM_{3,5}$ superconformal minimal model or $M_{4,5}$ minimal model in two dimensions \cite{Friedan:1984rv}. The truncation we study as a benchmark is 
\begin{align}
\phi \times \phi = 1 + [\phi^2] + [T] + [O_4]  ,
\end{align}
where $\Delta_{\phi}$ and $\Delta_4$ are parameters to be determined. The supersymmetry dictates that $\Delta_{\phi^2} = \Delta_{\phi} +1$ as in the supersymmetric Lee-Yang fixed point. The only difference is that there is no $\phi$ term in the right hand side due to the $\mathbb{Z}_2$ symmetry. In two dimensions, we know they are given by $\Delta_{\phi} = \frac{1}{5} $ and $\Delta_{4} = 4$.

We can study the prediction for $\Delta_{\phi}$ and $\Delta_4$  from the vanishing of $d_{124} = d_{234} = 0 $ by changing the space-time dimensions.
The results are shown in Fig \ref{fig:4-1} and Fig \ref{fig:4-2}. Again, there is no sign that the fixed points obtained by dimensional continuation disappear, and it predicts a non-trivial fixed point in three dimensions. We may further continue the dimensions until we reach $d=4$, where it is given by a free theory with conformal dimensions $\Delta_\phi=1.0$ and $\Delta_{4} = 6.0$. This is consistent with our understanding that the upper critical dimension of supersymmetric Ising model is four. We may further continue the space-time dimensions above four, but the significance of the formal (non-unitary)  solution in $d>4$ dimensions is not immediately obvious.


Let us make a small comment on the prediction in three dimensions. There has been some interest in finding conformal data of the supersymmetric Ising model in three dimensions. Our prediction from the truncated conformal bootstrap is $\Delta_{\phi} = 0.532$ and it is slightly smaller than the  predictions made by the other methods \cite{Fei:2016sgs}\cite{Gies:2017tod}\cite{Bashkirov:2013vya}\cite{Iliesiu:2015qra}\cite{Shimada:2015gda}\cite{Atanasov:2018kqw}. Most probably we are underestimating the conformal dimensions of $\phi$ as we can also see that it did underestimate in two dimensions; $\Delta_{\phi} = 0.144$ compared with the exact one $\Delta_{\phi} = \frac{1}{5}$. 

\section{Discussions}
In this paper, we have investigated an analogue of the supersymmetric Lee-Yang fixed point in three dimensions. The truncated conformal bootstrap seems to imply that its existence is not supported within the dimensional continuation.

Our findings do not necessarily exclude the existence of the fixed point that cannot be obtained by the dimensional continuation, but the definition of the supersymmetric Lee-Yang fixed point should be more accurately specified then. Another possibility is that the supersymmetric Lee-Yang fixed point does not preserve conformal symmetry, which is unlikely, but unless we give a precise definition of the supersymmetric Lee-Yang fixed point, we cannot  offer any argument for or against it.

In this paper, we have only focused on the single correlation function $\langle \phi(0) \phi(1) \phi(z) \phi(\infty) \rangle$, but given the supersymmetry we should obtain more constraint out of $\langle \phi^2(0) \phi^2(1) \phi^2(z) \phi^2(\infty) \rangle$ and $\langle \phi^2 (0) \phi^2(1) \phi(z) \phi(\infty) \rangle$. This may work well with the truncated conformal bootstrap approach because of the operator product expansion
\begin{align}
\phi^2 \times \phi^2 = 1 + [\phi] + [\phi^2] + [T] + \cdots \ ,
\end{align}
so we may be able to obtain twice number of constraints (by shifting the conformal dimension $\Delta_{\phi} \to \Delta_{\phi}+1$). This allows us to include more operators in the truncation, which might help improve the prediction.

Finally, in two dimensions, the supersymmetric Lee-Yang fixed point may be regarded as  a fermionic version of the non-unitary minimal model $M_{3,8}$. The dimensional continuation of $M_{3,8}$ does not have to preserve the supersymmetry and it may (or may not) give a different fixed point in three dimensions. A similar thing happens in the supersymmetric Ising model. If we regard it as the tri-critical Ising model, the dimensional continuation does not survive to three dimensions because the upper critical dimension of the tri-critical Ising model is three while we have the non-trivial supersymmetric Ising model in three dimensions. To address this issue, we need some more physical input about the nature of $M_{3,8}$ (e.g. Landau-Ginzburg effective action) that can be used in dimensional continuation, but this is still elusive.  

\section*{Acknowledgements}

This work is in part supported by JSPS KAKENHI Grant Number 17K14301.

\end{document}